\newif\ifshortver
\shortverfalse

\ifshortver
\documentclass[conference]{IEEEtran}
\else
\documentclass[english,onecolumn]{IEEEtran}
\fi
\usepackage[T1]{fontenc}
\usepackage[utf8]{inputenc}
\usepackage[cmex10]{amsmath}
\usepackage{amsthm}
\usepackage{amssymb}
\usepackage{graphicx}
\usepackage{esint}
\usepackage{amssymb,epsfig,graphics, graphicx, url,times,mathrsfs,algorithm,array,latexsym,fancyhdr,xspace,wrapfig, xcolor,multirow,amsthm,caption,cite,helvet,sidecap,bm,hyperref,url,subcaption}
\usepackage{bbm}
\makeatletter
\theoremstyle{plain}
\newtheorem{thm}{\protect\theoremname}
\theoremstyle{plain}
\newtheorem{conjecture}{\protect\conjecturename}
\theoremstyle{definition}

\theoremstyle{plain}
\newtheorem{claim}{\protect\claimname}

\usepackage{mathrsfs}
\usepackage{algorithm,algpseudocode}

\usepackage{colortbl}
\definecolor{lightgray}{rgb}{0.9,0.9,0.9}
\definecolor{lightred}{rgb}{1,0.8,0.8}
\definecolor{lightgreen}{rgb}{0.6,1,0.6}
\definecolor{lightyellow}{rgb}{1,1,0.5}
\definecolor{lightgrey}{rgb}{0.8,0.8,0.8}

\allowdisplaybreaks[1]

\makeatother

\usepackage{babel}
\providecommand{\conjecturename}{Conjecture}
\providecommand{\definitionname}{Definition}
\providecommand{\theoremname}{Theorem}
\providecommand{\claimname}{Claim}

\begin{document}
\title{Arithmetic Network Coding for Secret Sum Computation}
\author{Sijie Li and Cheuk Ting Li\\
Department of Information Engineering, The Chinese University of Hong
Kong\\
Email: sijieli@link.cuhk.edu.hk,  ctli@ie.cuhk.edu.hk}
\maketitle
\begin{abstract}
We consider a network coding problem where the destination wants to recover the sum of the signals (Gaussian random variables or random finite field elements) at all the source nodes, but the sum must be kept secret from an eavesdropper that can wiretap on a subset of edges.
This setting arises naturally in sensor networks and federated learning, where the secrecy of the sum of the signals (e.g. weights, gradients) may be desired.
While the case for finite field can be solved, the case for Gaussian random variables is surprisingly difficult.
We give a simple conjecture on the necessary and sufficient condition under which such secret computation is possible for the Gaussian case, and prove the conjecture when the number of wiretapped edges is at most 2.
\end{abstract}


\medskip{}

\section{Introduction}

Network coding, studied by Ahlswede \textit{et al.}~\cite{ahlswede2000network} and Li \textit{et al.}~\cite{li2003linear}, is a network communication technique where each node in the network combines its inputs to produce its outputs.
Network coding for sum computation, where the destination only wants to recover the sum of the sources, was investigated by and Ramamoorthy and Langberg~\cite{ramamoorthy2013communicating,langberg2009communicating}, Appuswamy \textit{et al.}~\cite{appuswamy2008network,appuswamy2009network,appuswamy2011network}, Shenvi and Dey~\cite{shenvi2010necessary}, and Rai and Dey~\cite{rai2012network}. Also see~\cite{rai2013capacity,tripathy2014sum,feizi2014network,guang2019improved}. This setting arises, for example, in sensor networks~\cite{rai2012network}.

This paper aims at solving the following network coding problem. Consider a network where each source node has a source signal that is a Gaussian random variable (or a uniformly chosen finite field element). The destination wants to recover the sum of the source signals, but the sum must be kept secret from an eavesdropper that can observe a subset of wiretapped edges in the network. There is no capacity constraint on the edges, which can transmit arbitrary signals (even real numbers as in arithmetic network coding~\cite{katti2007real,shintre2008real}). What is the condition under which this is possible?

In this paper, we resolve the case for finite field. For the Gaussian case, we present a simple conjecture on the necessary and sufficient condition under which such secret computation is possible -- secret computation is possible if and only if removing the wiretapped edges (and edges that are only reachable from the source nodes through wiretapped edges) does not weakly disconnect any source node from the destination node. Despite its simplicity, proving this conjecture appears to be surprisingly difficult. We prove the converse part of the conjecture, and the achievability part for the case where there are at most two wiretapped edges.

A similar problem has been studied by Guang \textit{et al.}~\cite{guang2021secure}, which considered secure sum computation over finite field where all sources (instead of only the sum) must be kept secret. We call our setting \emph{secret sum computation} to distinguish these two settings. While keeping all sources secret is desirable from a privacy perspective, there are settings where the secrecy of the sum is more important. For example, in a sensor network, the sensors are not users, so the privacy of individual sensors may be irrelevant. The sum (or average) of the measurements at the sensors is the quantity of interest, where decisions will be based upon, so its secrecy is vital.

Most aforementioned results on network coding for sum computation
are on finite fields or rings\footnote{We remark that arithmetic sum over integers has been studied in~\cite{appuswamy2009network,appuswamy2011network}. Nevertheless, the authors are not aware of any previous work on arithmetic sum over real numbers.}. Secrecy on sum of real numbers is considerably different from that for finite fields (e.g. the sum of two uniformly chosen elements in a finite field is independent of either one of the summands, but no analogous result holds for real numbers). We consider real source signals for two reasons. First, real signals is more natural for applications such as sensor networks and federated learning~\cite{mcmahan2017communication,kairouz2019advances}. Second, many existing protocols in federated learning and distributed averaging~\cite{xiao2004fast} involve sending real numbers, so it is not uncommon to allow transmission of real numbers. These will be elaborated in the following subsections.

\subsection{Federated Learning}

The proposed setting arises naturally in federated learning~\cite{mcmahan2017communication,kairouz2019advances}, a machine learning setting where the data is stored in a decentralized manner. The goal is to allow the server to train the model without having every user share all of their subsets of the data set. 
In federated stochastic gradient descent~\cite{shokri2015privacy}, each user train their local model using their subset of the data set, and upload the gradients of the parameters to the server. The server then update the global model according to the sum of the gradients. In federated averaging~\cite{mcmahan2017communication}, each user instead upload the parameters of their local model, and the server takes the average of those parameters. Both techniques involve sum computation (on the gradients or parameters) through a network. 

If a subset of the edges in the network is insecure and can be wiretapped, then it is desirable to ensure that the sum is independent of the signals along the wiretapped edges, so that the eavesdropper cannot gain any knowledge on the learned model at the server. We remark that this goal is different from ensuring the privacy of the local data sets of each user (which is more common in the literature), though for the real-valued signals case, ensuring that the eavesdropper's knowledge is independent of the sum can also guarantee privacy of each user to a certain degree (e.g., the eavesdropper cannot know the precise value of any individual source signal).

\subsection{Arithmetic Network Coding}

We do not impose any capacity constraint on the edges. Edges are allowed to carry arbitrary (even continuous) signals.
In the coding schemes constructed in this paper, real numbers are transmitted through the edges of the network, and only linear operations are performed at the nodes, which is the setting in (linear) arithmetic network coding~\cite{katti2007real,shintre2008real,nabaee2014quantized}. This is related to physical layer network coding~\cite{zhang2006hot,wilson2010joint,nazer2011reliable} and analog network coding~\cite{katti2007embracing,zhang2009optimal}, which are useful in wireless networks. Secure analog network coding has been studied, for example, in~\cite{mukherjee2010securing,kulhandjian2014securing}. 

Arithmetic network coding is a natural setting for federated learning, since many existing protocols in federated learning (e.g.~\cite{shokri2015privacy,mcmahan2017communication}) are based on sending real numbers. Also, several related protocols (e.g. distributed averaging~\cite{xiao2004fast}) involve transmitting real numbers. In practice, real numbers in these protocols are usually encoded as floating-point numbers, which are often accurate enough that the effect of quantization can be ignored. We also emphasize that the focus of this paper is not on the communication rate (where the encoding and quantization of those numbers would be of importance), but on the secrecy constraint.

Our linear coding scheme also applies to the physical layer network coding scenario where all edges are (single-use) additive Gaussian noise channels with noise variance $\sigma^2$ instead of having infinite capacities. In this case, a linear scheme can ensure secrecy of the sum, while allowing the sum to be recovered at the destination with mean squared error $O(\sigma^2)$. Refer to~\cite{hayashi2018secure} for a related setting in secure physical layer network coding with multiple access Gaussian channels. Also see~\cite{lu2009security,wang2011improving,jayasinghe2014secure,zhang2014beamforming} for other related results.

\subsection{Other Related Works}


The application of network coding on federated learning has been studied in~\cite{NCFL}, which shows that network coding can improve the communication efficiency. Nevertheless, \cite{NCFL} has only considered the butterfly network.

In the distributed averaging problem~\cite{xiao2004fast}, every node in the network has a number, and every node wants to obtain the average of these numbers. In our setting, only the server needs to obtain the sum (or average) of the numbers. Secure sum protocols~\cite{sheikh2009privacy,sheikh2016secure} are about computing the sum of the source signals while preserving the privacy of the source signal of each user. In our setting, we instead preserve the secrecy of the sum of the signals.

\medskip{}

This paper is organized as follows. In Section~\ref{sec:motivating}, we present a motivating example. In Section~\ref{sec:problem}, we define the problem setting. In Section~\ref{sec:finite}, we resolve the case for finite field signals. In Section~\ref{sec:gaussian}, we present the conjecture for the case for Gaussian signals, and prove the converse part of the conjecture, and the achievability part when the number of wiretapped edges is at most 2. 

\medskip{}

\section{Motivating Example}\label{sec:motivating}

\begin{figure}
\begin{centering}
\includegraphics[scale=0.575]{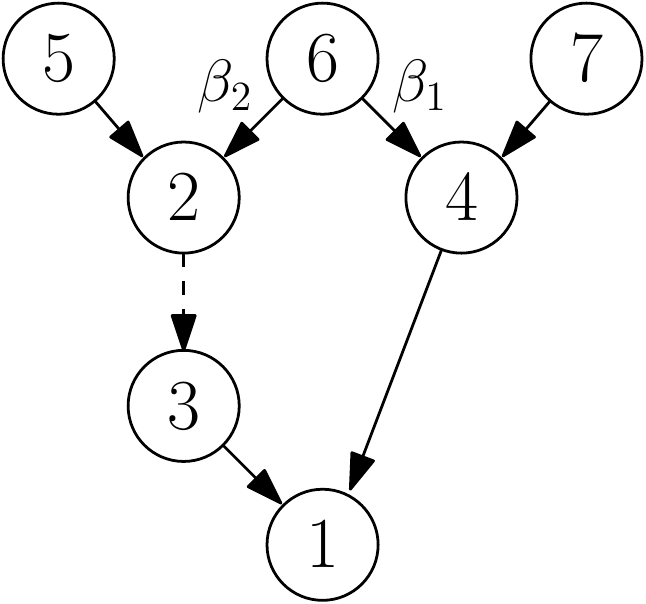}
\par\end{centering}
\caption{\label{fig:example}A simple network. The dashed edge is the wiretapped edge.}
\end{figure}

Consider the simple network in Figure~\ref{fig:example}, where each of the source nodes 5, 6, 7 holds an independent Gaussian source signal following $N(0,1)$ (or uniform in $\mathbb{F}_q$). Let the source signal at source node $s$ be $X_s$. The destination node $1$ wants to recover the sum $X_5+X_6+X_7$. Nevertheless, the signal along the wiretapped edge $(2,3)$ must be independent of the sum.

Consider a simple linear coding scheme where each of the source nodes 5, 7 forwards its source signal to its out-neighbor, and each of nodes 2, 3, 4 forwards the sum of its incoming signals to its out-neighbor. For source node $6$, it sends $\beta_1 X_6$ to node 4, and $\beta_2 X_6$ to node 2. The signal along edge $(2,3)$ is $X_5+ \beta_2 X_6$. For the Gaussian case, to ensure $X_5+ \beta_2 X_6$ is independent of $X_5+X_6+X_7$, we can take $\beta_2=-1$. For the finite field case, we can take any $\beta_2$. The sum of the signals to node 1 is $X_5 + X_7+(\beta_1 + \beta_2)X_6$. Taking $\beta_1=1-\beta_2$, we can ensure that the sum can be recovered.

To illustrate the difference between secret sum computation and secure sum computation,
note that if we require the signal along the wiretapped edge to be independent of $(X_5,X_6,X_7)$ (as in secure sum computation) instead of only the sum, then it is still possible if nodes are allowed to generate additional local randomness, since node 6 can generate a secret key and forward the key to node 4 (which in turn forwards it to node 1) and node 2, and node 2 can encrypt $X_5+X_6$ using the key. Node 1 can then recover $X_5+X_6$ using the key. Nevertheless, secrecy of $(X_5,X_6,X_7)$ is impossible if no local randomness is allowed. Secrecy of $(X_5,X_6,X_7)$ is also impossible for the Gaussian case if only linear codes are allowed (even if local randomness is allowed)\footnote{A secure sum computation scheme that ensures the secrecy of $(X_5,X_6,X_7)$ would be having node 6 generate $K \sim \mathrm{Unif}[0,1]$, and sending $K$ to node 4 (which in turn forwards it to node 1) and node 2. Node 2 sends $(\Phi((X_5+X_6)/\sqrt{2}) + K) \; \mathrm{mod} \; 1$, from which node 1 can recover $X_5+X_6$ using $K$, where $\Phi$ is the cdf of the standard Gaussian distribution. Nevertheless, this scheme is nonlinear and requires sending two numbers $X_7,K$ through edge $(4,1)$ (and hence fails when the edges are single-use additive Gaussian noise channels), and requires additional randomness $K$.}. In this paper, we require only the secrecy of the sum $X_5+X_6+X_7$, where linear codes suffice to show the achievability (for the case where there are at most two wiretapped edges), allowing the scheme to be also applicable to the case where all edges are (single-use) additive Gaussian noise channels. Also, in this paper, we do not assume the existence of local randomness.

\medskip{}

\section{Problem Formulation}\label{sec:problem}

A \emph{network} is defined as a weakly-connected directed acyclic
graph $(V,E)$, where we assume $V=\{1,\ldots,n\}$. For a node
$v$, denote its set of in-neighbors (resp. out-neighbors) as $N_{i}(v)$
(resp. $N_{o}(v)$). Nodes with zero in-degree are called \emph{source
nodes} (denote this set as $S:=\{v\in V:\,N_{i}(v)=\emptyset\}$),
and we require that there is exactly one node with zero out-degree
called the \emph{destination node} (let it be $d\in V$).

We consider the standard network coding setting, where each node can compute its outgoing signals as arbitrary functions of its incoming signals. The only difference is that there is no capacity constraint on the signals (they can be continuous). We give the precise definition of a \emph{secret coding scheme} as follows. First, each source node $s\in S$ observes an independent Gaussian source signal $X_{s}\sim N(0,1)$ (or a uniformly chosen element in the finite field $\mathbb{F}_q$)
i.i.d. across $s$. Then, in topological order, each node $v$ computes
its outgoing signals $Y_{v,w}$ along the edge $(v,w)$ for $w\in N_{o}(v)$
as a function of its observations (source $X_{v}$ if $v\in S$,
and signal $Y_{u,v}$ from in-neighbor $u\in N_{i}(v)$) without using any local randomness, i.e., 
\begin{equation}
Y_{v,w}:=f_{v,w}\big(\{Y_{u,v}\}_{u\in N_{i}(v)},\,\mathbbm{1}\{v\in S\}X_{v}\big)\label{eq:y_net}
\end{equation}
for $w\in N_{o}(v)$, where $\mathbbm{1}\{v\in S\} = 1$ if $v\in S$, $\mathbbm{1}\{v\in S\} = 0$ if $v\notin S$. Finally, the destination node $d$ computes
its output as
\begin{equation}
Z:= f_{d,\tilde{d}}\big(\{Y_{u,d}\}_{u\in N_{i}(d)}\big),\label{eq:z_net}
\end{equation}
where $\tilde{d}:=n+1$. The notation can be simplified by letting
$Y_{0,s}:=X_{s}$ for $s\in S$, and $Y_{d,\tilde{d}}:=Z$. Then,
instead of \eqref{eq:y_net} and \eqref{eq:z_net}, we have the following
formula: for any $v\in V$, $w\in N_{o}(v)$ or $(v,w)=(d,\tilde{d})$,
\[
Y_{v,w}=f_{v,w}\big(\{Y_{u,v}\}_{u\in N_{i}(v)\,\vee\,(u=0\,\wedge\,v\in S)}\big).
\]

Let $\tilde{E}\subseteq E$ be a set of \emph{wiretapped edges}, which
is accessible to the eavesdropper. A secret coding scheme with
wiretapped edges $\tilde{E}$ must satisfy two requirements. First, the
final output must be the sum of the sources, i.e.,
\[
Y_{d,\tilde{d}}=\sum_{s\in S}Y_{0,s}.
\]
Second, the collection of signals along the edges $\tilde{E}$ must be independent
of the sum, i.e.,
\[
\{Y_{u,v}\}_{(u,v)\in\tilde{E}}\perp\!\!\!\perp \sum_{s\in S}Y_{0,s}.
\]
This ensures that the eavesdropper cannot gain any information about
the sum.

We may also consider a \emph{linear secret coding scheme}, where
\[
f_{v,w}\big(\{Y_{u,v}\}_{u\in N_{i}(v)\,\vee\,(u=0\,\wedge\,v\in S)}\big) = \sum_{u\in N_{i}(v)\,\vee\,(u=0\,\wedge\,v\in S)}\alpha_{u,v,w}Y_{u,v}
\]
is a linear combination of its inputs as in arithmetic network coding~\cite{katti2007real,shintre2008real}, where $\{\alpha_{u,v,w}\}$ is a collection of coefficients. While our conjecture and the converse result are on general (linear/nonlinear) secret coding schemes, the schemes we construct in achievability proofs are linear secret coding schemes.

We use the notation
$u \longrightarrow v$ to mean that $v$ is reachable from $u$, i.e., there is a directed path from $u$ to $v$.
The notation $u \overset{\tilde{E}^c}{\longrightarrow} v$ means that there is a directed path from $u$ to $v$ in the graph where the edges in $\tilde{E}$ are removed. 

\section{Finite Field Source Signals}\label{sec:finite}

In this section, we resolve the case for source signals in the finite field $\mathbb{F}_q$.

\begin{thm}
Consider the network $(V,E)$ with uniform signals in $\mathbb{F}_q$ and wiretapped edges $\tilde{E}\subseteq E$. A secret coding
scheme exists if and only if there exists $s \in S$ such that $s \overset{\tilde{E}^c}{\longrightarrow} d$.
\end{thm}
\begin{IEEEproof}
For the converse part, assume removing the edges in $\tilde{E}$ disconnects all source nodes from the destination. Then the collection of source signals $\{Y_{0,s}\}_{s\in S}$ is conditionally independent of the output of the destination $Y_{d,\tilde{d}}$ (which should be the sum of sources), conditioned on the collection of the signals along the wiretapped edges $\{Y_{u,v}\}_{(u,v)\in\tilde{E}}$ (which should be independent of the sum of sources). This is clearly impossible.

For the achievability part, let $s_0 \in S$ with $s_0 \overset{\tilde{E}^c}{\longrightarrow} d$. Consider the path $v_0=s_0,v_1,\ldots,v_{k-1},v_k=d$ from $s_0$ to $d$ without passing through $\tilde{E}$. Design a linear secret coding scheme as follows. Each node $u$ forwards the sum of its inputs (and its source signal if it is a source node) to one of its out-neighbors, where we choose the out-neighbor $v_{i+1}$ if $u=v_i$ for some $i$, or any one out-neighbor of $u$ if $u$ is not on the path. Note that only the edges on the path can depend on the source signal $Y_{0,s_0}$. Each edge not on the path has a signal that is the sum of a subset in $\{Y_{0,s}\}_{s\in S \backslash \{s_0\}}$. Hence, $\{Y_{u,v}\}_{(u,v)\notin \{(v_0,v_1),\ldots,(v_{k-1},v_k)\}}$ (the collection of signals along edges not on the path) is a function of $\{Y_{0,s}\}_{s\in S \backslash \{s_0\}}$, which is independent of the sum $\sum_{s\in S}Y_{0,s}$.
\end{IEEEproof}

The finite field case is simple and perhaps not particularly enlightening, considering that the sum can be kept secret even when all except one of the sources are leaked, making such scheme useful only when the privacy of individual sources is unimportant. On the contrary, for the Gaussian case, keeping the sum secret implies that the eavesdropper cannot know any of the sources exactly, making it a harder requirement to fulfill.

\medskip

\section{Gaussian Source Signals}\label{sec:gaussian}


In the remainder of the paper, we consider the case for Gaussian source signals.
We first make an observation that if the set of edges $\tilde{E}\subseteq E$ has signals that are independent of the sum of the sources, then any edge that cannot be visited from a source node without passing through $\tilde{E}$ must also be independent of the sum. Define the \emph{secrecy closure} of $\tilde{E}$ as
\[
\mathrm{scl}(\tilde{E}) := \tilde{E} \cup \{(u,v) \in E:\, \lnot \exists s \in S.\, s \overset{\tilde{E}^c}{\longrightarrow} u\}.
\]
Since the signals along the edges in $\mathrm{scl}(\tilde{E})$ only depend on the signals along the edges in $\tilde{E}$, the edges in $\mathrm{scl}(\tilde{E})$ must also be independent of the sum. 

We now present our main conjecture on the necessary and sufficient condition for a secret coding scheme to exist.
\begin{conjecture}\label{conj:main}
Consider the network $(V,E)$ with Gaussian source signals and wiretapped edges $\tilde{E}\subseteq E$. A secret coding
scheme exists if and only if removing
the edges in $\mathrm{scl}(\tilde{E})$ does not weakly disconnect any source node
$s\in S$ from the destination node $d$ (i.e., in the graph $(V,\, E\backslash\mathrm{scl}(\tilde{E}))$, after treating the graph as an undirected graph,
any $s\in S$ is connected to $d$).
\end{conjecture}

\ifshortver
\else
We first show the converse part of the conjecture. The proof is given in Appendix~\ref{subsec:pf_converse}.
\fi
\begin{thm} [Converse]
\label{thm:converse}Consider the network $(V,E)$  with Gaussian source signals and wiretapped edges $\tilde{E}\subseteq E$. If removing
the edges in $\mathrm{scl}(\tilde{E})$  weakly disconnect a source node
$s\in S$ from the destination node $d$, then there does not exist a secret coding
scheme with wiretapped edges $\tilde{E}$.
\end{thm}




Next, we discuss the achievability part of the conjecture. Despite the simplicity of the conjecture, proving the achievability is surprisingly hard, and we are only able to prove it for $|\tilde{E}| \le 2$.  We first give a sketch of the proof of the case where there is one wiretapped edge. 
\ifshortver
\else
Refer to Theorem~\ref{thm:twoedges} 
\fi
for the rigorous proof (note that the two-edge case covers the one-edge case since we can split the one wiretapped edge into two).

\begin{thm} [One wiretapped edge] \label{thm:oneedge}
Consider the network $(V,E)$ with Gaussian source signals and wiretapped edges $\tilde{E}\subseteq E$ with $|\tilde{E}|=1$. A linear secret coding
scheme exists if removing
the edges in $\mathrm{scl}(\tilde{E})$ does not weakly disconnect any source node
$s\in S$ from the destination node $d$.
\end{thm}
\begin{IEEEproof}[Proof sketch]
Without loss of generality, assume $\tilde{E} = \{(2, v_2)\}$ and $d=1$. We assign label to each node $u$ according to which nodes in node $1,2$ that $u$ can reach without going through $\tilde{E}$, that is,
\begin{align*}
l(u)&=\begin{cases}
1 & \mathrm{if}\;u \overset{\tilde{E}^c}{\longrightarrow} 1,\; u \overset{\tilde{E}^c}{\not\longrightarrow} 2\\
2 & \mathrm{if}\;u \overset{\tilde{E}^c}{\not\longrightarrow} 1,\; u \overset{\tilde{E}^c}{\longrightarrow} 2\\
12 & \mathrm{if}\;u \overset{\tilde{E}^c}{\longrightarrow} 1,\; u \overset{\tilde{E}^c}{\longrightarrow} 2.
\end{cases}
\end{align*}
Note that any node falls in one of the above three cases (if removing edge $(2,v_2)$ disconnects a node from the destination, then that node must be able to reach $2$). Also note that $l(v_2)=l(v_3)=1$.

Define $S_{\lambda} := \{u \in S :\, l(u)=\lambda\}$ to be the set of source nodes with label $\lambda$, and $V_{\lambda} := \{u \in V :\, l(u)=\lambda\}$. Note that $S_2 \cup S_{12} \neq \emptyset$ since we can find an ancestor of node $2$ with zero in-degree, which must be a source node. Fix any $s \in S_2\cup S_{12}$. Since removing the edge $(2,v_2)$ does not weakly connect $s$ from node 1, we can consider an undirected path from $s$ to node 1. Since $l(s) \in \{2,12\}$ and $l(1)=1$, there must be an edge $(w_1,w_2)\in E \backslash \tilde{E}$ along that path that connects two nodes $w_1,w_2$ with different labels. Note that $l(w_1)=12$ since at least one of $w_1$ and $w_2$ must be able to reach node 1 (and the same holds for node 2). By considering an ancestor of $w_1$ with zero in-degree, we know that $S_{12} \neq \emptyset$.

\begin{figure}
\begin{centering}
\includegraphics[scale=0.57]{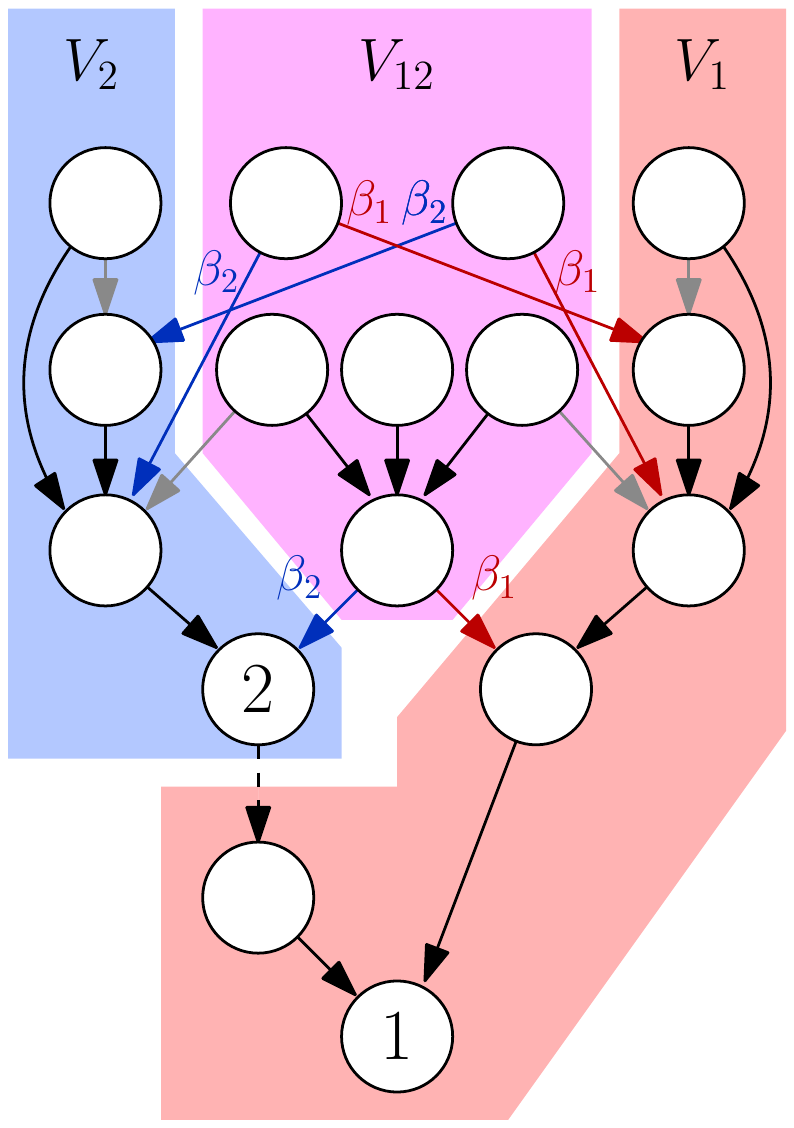}
\par\end{centering}
\caption{\label{fig:oneedge}The coding scheme in Theorem~\ref{thm:oneedge}. The dashed edge is the wiretapped edge. The nodes are partitioned into $V_1,V_2,V_{12}$ according to which of nodes $1,2$ it can reach. There are 4 kinds of edge: black edges which forward the sum of the incoming signals, red edges which forward the sum after multiplying by $\beta_1$, blue edges which forward the sum after multiplying by $\beta_2$, and grey edges which are unused.}
\end{figure}

We now design a linear secret coding scheme. For each node $u \notin \{1,2\}$, if it has an out-neighbor $v$ (via a non-wiretapped edge) with the same label, then node $u$ forwards the sum of its incoming signals to node $v$. If there are multiple possible $v$'s, select any one of them. If there is no such $v$, then $l(u)=12$ must hold, and node $u$ will instead forward the sum of its incoming signals, after multiplying by $\beta_{1}$, to one of its out-neighbors with label 1; and also forward the sum of its incoming signals, after multiplying by $\beta_{2}$, to one of its out-neighbors with label 2. For node 2, it simply forward the sum of its incoming signals along the wiretapped edge. Refer to Figure~\ref{fig:oneedge} for an illustration.

Write $X_{A} = \sum_{v \in A} X_{v}$. The signal along the wiretapped edge $(2,v_2)$ is
\[
Y_{2,v_2} = X_{S_2} + \beta_{2} X_{S_{12}}.
\]
Since the signals are Gaussian, to check the secrecy constraint, it suffices to check that the covariance
\[
\mathbf{E}[(X_{S_1}+X_{S_2}+X_{S_{12}})Y_{2,v_2}] = |S_2| + \beta_{2}|S_{12}| = 0,
\]
which gives $\beta_{2} = -|S_2|/|S_{12}|$. The sum of the signals at the destination node $1$ is
\[
X_{S_1} + \beta_{1} X_{S_{12}} + Y_{2,v_2} = X_{S_1} + X_{S_2} + (\beta_{1}+\beta_{2}) X_{S_{12}}.
\]
For the recovery requirement, we take $\beta_{1}=1-\beta_{2}=1+|S_2|/|S_{12}|$.
\end{IEEEproof}

\medskip

We now discuss the case with two wiretapped edges.

\begin{thm} [Two wiretapped edges]\label{thm:twoedges}
Consider the network $(V,E)$ with Gaussian source signals and wiretapped edges $\tilde{E}\subseteq E$ with $|\tilde{E}|=2$. A linear secret coding 
scheme exists if removing
the edges in $\mathrm{scl}(\tilde{E})$ does not weakly disconnect any source node
$s\in S$ from the destination node $d$.
\end{thm}

\begin{figure}
\begin{centering}
\includegraphics[scale=0.7]{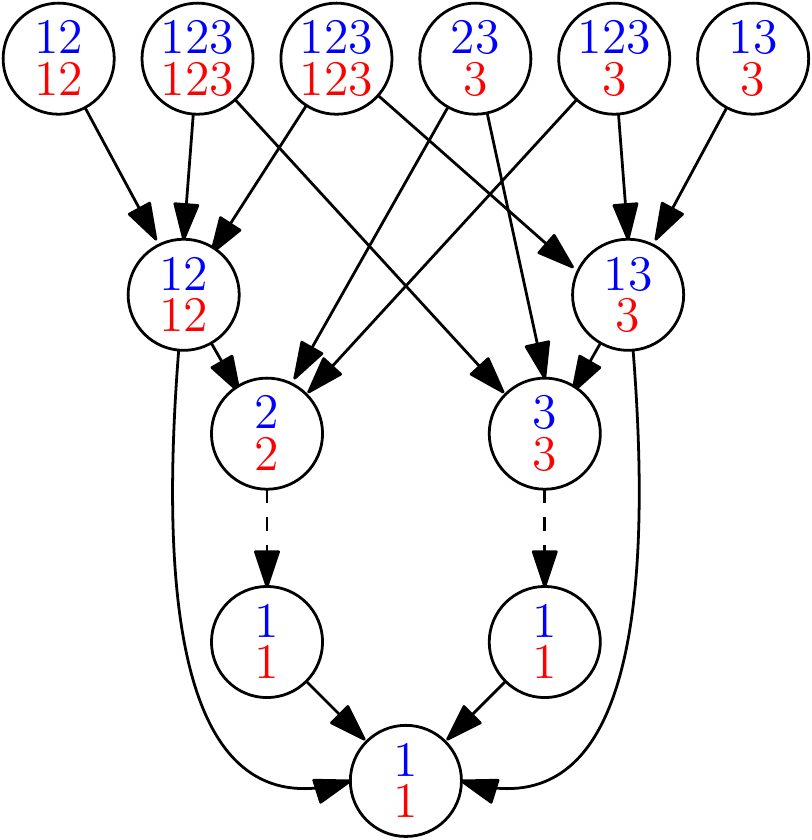}
\par\end{centering}
\caption{\label{fig:twoedge}An example where there exist source nodes with label $123$. The original labels $l(u)$ are in blue. The proof involves assigning new labels $l'(u)$ to each node (in red, by the ``relabelling to $3$'' rule).}
\end{figure}

\begin{IEEEproof}[Sketch of the proof of Theorem \ref{thm:twoedges}]
Assume $d=1$ and the wiretapped edges are $(2,v_2)$ and $(3,v_3)$. A natural attempt is to define a labelling according to which of nodes 1, 2, 3 can the node reach as in the proof Theorem~\ref{thm:oneedge}, resulting in seven labels: 1, 2, 3, 12, 13, 23, 123.
For the case where there is no source node with label $123$, it can be solved using a similar technique as Theorem~\ref{thm:oneedge}.
However, for the case where there is a source node with label $123$, there are many different cases of the set of labels of its out-neighbors (it can be $\{1, 23\}$, $\{12, 13\}$, etc.), and hence we cannot define coefficients for nodes with label 123 together. We should rather treat $\{1, 23\}$, $\{12, 13\}$, etc. as separate labels. But now a node with set of labels of its out-neighbors $\{\{1, 23\}, \{12, 13\}\}$ needs to have its own label as well, resulting in an unbounded number of labels. 

To reduce the number of labels, we consider the following new labelling rule, which we call \emph{relabelling to $3$}. Nodes with labels $1$, $2$, $3$ and $12$ remain unchanged. If a node has out-neighbors with labels $1$ and $3$ only, then it is labelled $3$ instead of $13$. If a node has out-neighbors with labels $2$ and $3$ only, then it is labelled $3$ instead of $23$. If a node has out-neighbors with labels $12$ and $3$, then it is labelled $123$. There are five different labels: $1$, $2$, $3$, $12$, $123$. Now, a node with label $123$ either has out-neighbors with labels $12$ and $3$, or an out-neighbor with label $123$, so we only have to design the coefficients to labels $12$ and $3$ (instead of considering $\{1, 23\}, \{12, 13\}$ etc.). Refer to Figure~\ref{fig:twoedge} for an illustration of the original and new labelling rules.

Nevertheless, it is possible that the information of the weakly-connectedness assumption in Theorem \ref{thm:twoedges} is lost in the new labelling rule. For example, if the only node with original label $123$ is a source node with out-neighbors with original labels $13$ and $2$, then the weakly-connectedness assumption is satisfied. Nevertheless, this source node will have a new label $3$ (since its out-neighbors have new labels $3$ and $2$). Hence, if we only consider the new labels, then it is possible that we only have source nodes with labels $1$, $2$ and $3$ even if weakly-connectedness is satisfied. Therefore, it is impossible to deduce whether weakly-connectedness is satisfied using only the labels of the source node. A scheme that only depends on the new labels of the node must fail due to the converse of the conjecture.

To solve this problem, we apply the new labelling rule according to the out-neighbors of any one node with original label $123$. If the out-neighbors have labels $13$ and $2$, then we apply a relabelling to $2$ instead of $3$ (i.e., we have label $2$ instead of labels $12$, $23$). If the out-neighbors have labels $23$ and $1$, then we apply a relabelling to $1$. This way, we can guarantee that this node with original label $123$ will still have a new label $123$. The presence of a source node with label $123$ implies that weakly-connectedness is satisfied. Hence, we can design the coefficients for the new labels in a way similar to Theorem~\ref{thm:oneedge}.

\end{IEEEproof}

\ifshortver
\else
\medskip

We now give the rigorous proof.

\begin{IEEEproof}[Proof of Theorem \ref{thm:twoedges}]
Define 
\[
\mathrm{vscl}(\tilde{E}) :=  \{u \in V:\, \lnot \exists s \in S.\, s \overset{\tilde{E}^c}{\longrightarrow} u\}.
\]
to be the set of nodes only reachable from the source nodes via $\tilde{E}$. Then we have $\mathrm{scl}(\tilde{E}) = \tilde{E} \cup \{(u,v) \in E:\, u \in \mathrm{vscl}(\tilde{E})\}$.

Let $\tilde{E} = \{(u_2,v_2), (u_3,v_3)\}$.
Without loss of generality, assume $u_2 = 2$, $u_3 = 3$ and $d=1$.
We will assign label $l(u) \subseteq \{1,2,3\}$ to node $u$ according to the following rule:
\[
l(u) := \{v \in \{1,2,3\}:\, u \overset{\tilde{E}^c}{\longrightarrow} v\}.
\]
Define 
\begin{align*}
    V_{\lambda} &:= \{u \in V :\, l(u)=\lambda\}, \\
    S_{\lambda} &:= \{u \in S :\, l(u)=\lambda\}.
\end{align*}

We write $(V,\tilde{E}^c)$ for the graph where the edges in $\tilde{E}$ are removed.
Note that the labels $l(u)$ are subsets of $\{1,2,3\}$, and we will write $12$ for the subset $\{1,2\}$ for brevity. Also note that $l(u)\neq \emptyset$ since we have $u \longrightarrow 1$ in the original graph (node $1$ is the only node with zero out-degree), so after removing the edges $(2,v_1), (3,v_2)$, we have $u \overset{\tilde{E}^c}{\longrightarrow} 1$, $u \overset{\tilde{E}^c}{\longrightarrow} 2$ or $u \overset{\tilde{E}^c}{\longrightarrow} 3$.

Moreover, we assume that $2,3 \notin \mathrm{vscl}(\tilde{E})$. If $3 \in \mathrm{vscl}(\tilde{E})$, then node $3$ is only reachable from source nodes via the wiretapped edge $(2,v_2)$, so the secrecy requirement on $\{(2,v_2),(3,v_3)\}$ is implied by the secrecy requirement on $(2,v_2)$, and this reduces to the one-wiretapped-edge case. Hence we can assume $2,3 \notin \mathrm{vscl}(\tilde{E})$.

Also, we assume that $l(2)=\{2\}$ and $l(3)=\{3\}$. If $l(2)\neq \{2\}$, then we can add a new node $2'$, remove the edge $(2,v_2)$, and add the edges $(2,2')$ and $(2',v_2)$ (which is the new wiretapped edge). Adding a node this way is permitted since node $2$ in the original network can simulate node $2'$ in the new network. Now we have $l(2')=\{2\}$, and hence we can use node $2'$ as the new node $2$. Therefore, we can always assume $l(2)=\{2\}$ and $l(3)=\{3\}$.

Note that at least one of $l(v_2)$, $l(v_3)$ is $\{1\}$ (otherwise it forms a cycle). Without loss of generality, assume $l(v_2)=\{1\}$. We have $l(v_3)=\{1\}$, $\{2\}$ or $\{1,2\}$.

We first prove the following claims:

\begin{claim}\label{claim:s_scl}
If $s \in S$, $v \in V$, $s \overset{\tilde{E}^c}{\longrightarrow} v$, then $s \overset{\mathrm{scl}(\tilde{E})^c}{\longrightarrow} v$.
\end{claim}
To prove this claim, assume the contrary that $s \overset{\tilde{E}^c}{\longrightarrow} v$, $s \overset{\mathrm{scl}(\tilde{E})^c}{\not\longrightarrow} v$. Then there exists a directed path from $s$ to $v$ in $(V,\tilde{E}^c)$ that passes through an edge in $\mathrm{scl}(\tilde{E}) \backslash \tilde{E}$, contradicting the definition of $\mathrm{scl}$.
\medskip

\begin{claim}\label{claim:vne_imp_sne}
For any node $v \notin \mathrm{vscl}(\tilde{E})$ with $l(v) \neq \{1\}$,
there exists $\beta \supseteq l(v)$ such that $S_{\beta} \neq \emptyset$.
\end{claim}
To prove this claim, let $v \notin \mathrm{vscl}(\tilde{E})$, $l(v)=\lambda$, $\lambda \neq \{1\}$. 
By the definition of $\mathrm{vscl}$, there is an ancestor $s$ of $v$ in the graph $(V,\tilde{E}^c)$ that is a source node. We have $l(s) \supseteq l(v)$, and the claim is satisfied.

\medskip

\begin{claim}\label{s123}
If $S_{123}= \emptyset$, then at least two of $S_{12}$, $S_{13}$ and $S_{23}$ are nonempty.
\end{claim}
To prove the claim, assume $S_{123} = \emptyset$. Since node $1$ is weakly connected to node $2$ in the graph $(V,\mathrm{scl}(\tilde{E})^c)$ (by Claim \ref{claim:vne_imp_sne}, since $2 \notin \mathrm{vscl}(\tilde{E})$, there exists $s\in S$ where $s \overset{\tilde{E}^c}{\longrightarrow} 2$, and hence $s \overset{\mathrm{scl}(\tilde{E})^c}{\longrightarrow} 2$ by Claim~\ref{claim:s_scl}, and $s$ is weakly connected to $1$ in $(V,\mathrm{scl}(\tilde{E})^c)$ by the assumption in the theorem), 
we can find a sequence of nodes $w_0 = 1, w_1, \ldots ,w_{k-1}, w_k = 2$ such that any two consecutive nodes $w_{i-1}, w_i$ are adjacent in $(V,\mathrm{scl}(\tilde{E})^c)$ (i.e., either $(w_{i-1}, w_i) \in E \backslash \mathrm{scl}(\tilde{E})$ or  $(w_i, w_{i-1}) \in E \backslash \mathrm{scl}(\tilde{E})$). Consider $l(w_{i-1}) \cup l(w_i)$. 
By Claim \ref{claim:vne_imp_sne}, if $l(w_{i-1}) \cup l(w_i) = \lambda$, $|\lambda| = 2$, then we have $S_{\lambda} \neq \emptyset$ (this is because if $(w_{i-1}, w_i) \in E \backslash \mathrm{scl}(\tilde{E})$, then $l(w_{i-1}) = \lambda$ and $w_{i-1} \notin \mathrm{vscl}(\tilde{E})$ by the definition of $\mathrm{scl}$). 
If there exists $i$ such that $l(w_{i-1}) \cup l(w_i) = \{1,2\}$, then $S_{12} \neq \emptyset$. If there does not exist such $i$, then let $i_{\min}, i_{\max}$ be the minimum / maximum $i$ such that $|l(w_{i-1}) \cup l(w_i)| \ge 2$ (such $i$ exists since $1 \in l(w_0)$ and $2 \in l(w_k)$). Since $\{1,2\} \nsubseteq l(w_{i-1}) \cup l(w_i)$, we have $l(w_{i_{\min}-1}) \cup l(w_{i_{\min}}) = \{1, 3\}$, implying $S_{13}\neq \emptyset$. Similarly, by considering $i_{\max}$, we have $S_{23}\neq \emptyset$. Hence, we have $S_{12} \neq \emptyset$, or $S_{13},S_{23} \neq \emptyset$. By considering nodes 1 and 3 instead of nodes 1 and 2, we can similarly show that $S_{13} \neq \emptyset$, or $S_{12},S_{23} \neq \emptyset$. Combining these two conditions, we know that if $S_{123}= \emptyset$, then at least two of $S_{12}$, $S_{13}$ and $S_{23}$ are nonempty.

\medskip

Then we have two cases on whether $S_{123} \neq \emptyset$ or not. We will design the coefficients in the scheme for the two cases separately. 

\medskip


\textbf{Case 1:} 
Suppose $S_{123} = \emptyset$. By Claim \ref{s123}, at least two sets of $S_{12}$, $S_{13}$ and $S_{23}$ are nonempty. 
We design the scheme as follows. For a node $v \in V_{12}$, it forwards the sum of its incoming signals to one of its out-neighbors $w \in V_{12}$ (along a non-wiretapped edge) with the same label. If no such out-neighbor exists, then it forwards the sum of its incoming signals to one of its out-neighbors $w_1 \in V_1$ (along a non-wiretapped edge) after multiplying the sum by the coefficient $\beta_{12, 1}$, and also forwards the sum to one of its out-neighbors $w_2 \in V_2$ (along a non-wiretapped edge) after multiplying the sum by the coefficient $\beta_{12, 2}$, where $\{\beta_{\lambda_1, \lambda_2}\}_{\lambda_2 \subseteq \lambda_1 \subseteq\{1,2,3\}}$ is a set of coefficients. Perform similar operations for $V_{13},V_{2,3}$. More generally, for $v \notin \{1,2,3\}$, we use a linear secret coding scheme with
\begin{align}
\alpha_{u,v,w}&=\begin{cases}
\mathbbm{1}\{w=\min(\tilde{N}_{o}(v)\cap V_{l(v)})\} & \mathrm{if}\;\tilde{N}_{o}(v)\cap V_{l(v)}\neq\emptyset\\
\beta_{l(v),l(w)}\mathbbm{1}\{w=\min(\tilde{N}_{o}(v)\cap V_{l(w)})\} & \mathrm{if}\;\tilde{N}_{o}(v)\cap V_{l(v)}=\emptyset,
\end{cases}\label{eqn:alpha}
\end{align}
where $\tilde{N}_{o}(v)$ is the set of out-neighbors of node $v$ in graph $\tilde{E}^{c}$. Note that the ``$w=\min(\cdots)$'' condition is to ensure that the signal is only forwarded to one out-neighbor in the first case, or two out-neighbors in the second case.

For node $2$ or $3$, it simply forwards the sum of its incoming signals through the wiretapped edge. More precisely, we take $\alpha_{u,2,v_2} = 1$ for any $u$, and $\alpha_{u,3,v_3} = 1$ for any $u$.

First consider the case $l(v_3)=\{1\}$. Note that for each source node in $S_1$, it will reach node $1$ without being multiplied by any coefficient (i.e., the $\alpha_{u,v,w}$'s along its path to node $1$ will be all $1$'s). The same for source nodes in $S_2$ and $S_3$ (since $\alpha_{u,v,w} = 1$ along the wiretapped edge). For a source node in $S_{12}$, it will encounter the second case in \eqref{eqn:alpha} exactly once, and be forwarded to two nodes in $V_1$ and $V_2$ respectively after being multiplied by $\beta_{12,1}$ and $\beta_{12,2}$ respectively. Hence, it will reach node $2$ after being multiplied by $\beta_{12,2}$, and reach node $1$ after being multiplied by $\beta_{12,1} + \beta_{12,2}$. Similar for $S_{13},S_{23}$. Write $X_{A} = \sum_{v \in A} X_{v}$. The signals along the wiretapped edges $(2,v_2)$, $(3,v_3)$ are
\begin{align}
Y_{(2,v_2)} & = X_{S_2} + \beta_{12,2} X_{S_{12}} + \beta_{23,2} X_{S_{23}},\label{eqn:y2}\\
Y_{(3,v_3)} & = X_{S_3} + \beta_{13,3} X_{S_{13}} + \beta_{23,3} X_{S_{23}},\label{eqn:y3}
\end{align}
and the sum of the signals to node $1$ is
\begin{align}
& X_{S_1} + \beta_{12,1} X_{S_{12}} + \beta_{13,1} X_{S_{13}} + Y_{(2,v_2)} + Y_{(3,v_3)} \nonumber \\
& = X_{S_1} + X_{S_2} + X_{S_3} + (\beta_{12,1} + \beta_{12,2})X_{S_{12}} + (\beta_{13,1}+\beta_{13,3}) X_{S_{13}} + (\beta_{23,2}+\beta_{23,3}) X_{S_{23}}.\label{eqn:node1}
\end{align}
For the secrecy constraint, we require that $Y_{(2,v_2)}$ and $Y_{(3,v_3)}$ are uncorrelated with the sum of $X_s$'s. By \eqref{eqn:y2} and \eqref{eqn:y3}, we have
\begin{align}
|S_2| + \beta_{12,2} |S_{12}| + \beta_{23,2} |S_{23}| = 0,\nonumber\\
|S_3| + \beta_{13,3} |S_{13}| + \beta_{23,3} |S_{23}| = 0.\label{eqn:sy3}
\end{align}
For the recovery requirement, we need \eqref{eqn:node1} to be the sum of $X_s$'s. Hence,
\begin{align}
\beta_{12,1} + \beta_{12,2} &= 1,\nonumber\\
\beta_{13,1}+\beta_{13,3} &= 1,\nonumber\\
\beta_{23,2}+\beta_{23,3} &= 1.\label{eqn:recovery}
\end{align}

Next consider the case $l(v_3)=\{1,2\}$. The only difference is that now the signal along the wiretapped edge $(3,v_3)$ is sent to a node with label $\{1,2\}$.
The signals along the wiretapped edges $(3,v_3)$, $(2,v_2)$ are
\begin{align*}
Y_{(3,v_3)} & = X_{S_3} + \beta_{13,3} X_{S_{13}} + \beta_{23,3} X_{S_{23}},\\
Y_{(2,v_2)} & = X_{S_2} + \beta_{12,2} (X_{S_{12}}+Y_{(3,v_3)}) + \beta_{23,2} X_{S_{23}}\\
& = X_{S_2}+ \beta_{12,2} X_{S_3} + \beta_{12,2} X_{S_{12}} + \beta_{12,2}\beta_{13,3} X_{S_{13}} \\
& \;\;\;\; + (\beta_{12,2}\beta_{23,3} + \beta_{23,2}) X_{S_{23}},
\end{align*}
and the sum of the signals to node $1$ is
\begin{align*}
& X_{S_1} + \beta_{12,1} (X_{S_{12}} + Y_{(3,v_3)}) + \beta_{13,1} X_{S_{13}} + Y_{(2,v_2)}  \\
& = X_{S_1} + X_{S_2} + (\beta_{12,1} + \beta_{12,2}) (X_{S_3}+ X_{S_{12}}) \\
&\;\;\;\; + ((\beta_{12,1}+\beta_{12,2})\beta_{13,3} + \beta_{13,1}) X_{S_{13}} \\
& \;\;\;\; +  ((\beta_{12,1}+\beta_{12,2})\beta_{23,3} + \beta_{23,2}) X_{S_{23}}.
\end{align*}
It is straightforward to check that the secrecy constraint and the recovery requirement reduce to exactly the same constraints~\eqref{eqn:sy3}, \eqref{eqn:recovery} as in the case $l(v_3)=\{1\}$. The intuitive reason is that, since both sources in $S_1$ or $S_{12}$ would be transmitted to the destination with coefficient $1$, whether $l(v_3)=\{1\}$ or $l(v_3)=\{1,2\}$ would not change whether the recovery requirement is satisfied. Also, since the signal along the wiretapped edge $(3,v_3)$ is uncorrelated to the sum, adding that signal to a node with label $\{1,2\}$ would not change whether the secrecy constraint at edge $(2,v_2)$ is satisfied.
The case $l(v_3)=\{2\}$ is the same as the case $l(v_3)=\{1\}$ for the same reason.

Recall that at least two of $S_{12}$, $S_{13}$ and $S_{23}$ are nonempty. We consider the following cases.

\textbf{Case 1a:} $S_{12}, S_{23} \neq \emptyset$. We can take
\begin{equation}
    \left \{        
    \begin{array}{cl}
    \beta_{12,1} &=  1-\beta_{12,2}  \\
    \beta_{12,2} &= -\frac{|S_{2}|+\beta_{23,2}|S_{23}|}{|S_{12}|}   \\
    \beta_{23,2} &= 1- \beta_{23,3}   \\
    \beta_{23,3} &= -\frac{|S_3|}{|S_{23}|}  \\
    \beta_{13,1} &= 1 \\
    \beta_{13,3} &= 0. \\
    \end{array}
 \right.
\end{equation}

\textbf{Case 1b:} $S_{12}, S_{13} \neq \emptyset$, $S_{23} = \emptyset$, We can take
\begin{equation}
    \left \{
    \begin{array}{cl}
        \beta_{12,1} &= 1-\beta_{12,2}   \\
        \beta_{12,2} &= -\frac{|S_{2}|}{|S_{12}|}  \\
        \beta_{13,1} &= 1-\beta_{13,3}  \\
        \beta_{13,3} &= -\frac{|S_{3}|}{|S_{13}|} \\
        \beta_{23,2} &= 1   \\
        \beta_{23,3} &= 0.  \\
    \end{array}
    \right.
\end{equation}
The case $S_{13}, S_{23} \neq \emptyset$ is similar.

\medskip

\textbf{Case 2:}
Suppose $S_{123} \neq \emptyset$.
Fix any $s \in S_{123}$. Consider the case where there does not exist node $v$ with $|l(v)|=2$ such that $s \overset{\tilde{E}^c}{\longrightarrow} v$. Let $w$ be a node satisfying $l(w)=\{1,2,3\}$, $s \overset{\tilde{E}^c}{\longrightarrow} w$, and there is no out-neighbor of $w$ with label $\{1,2,3\}$ (such $w$ can be found by starting at $s$ and repeatedly moving to an out-neighbor of the current node while staying at nodes with label $\{1,2,3\}$). Due to the assumption that there does not exist node $v$ with $|l(v)|=2$ such that $s \overset{\tilde{E}^c}{\longrightarrow} v$, there exists out-neighbors $w_1,w_2,w_3$ of $w$ where $l(w_1)=\{1\}$, $l(w_2)=\{2\}$ and $l(w_3)=\{3\}$. We can create a new node $w'$, remove edges $(w,w_1),(w,w_2)$, and add edges $(w,w'),(w',w_1),(w',w_2)$. Adding a node this way is permitted since node $w$ in the original network can simulate node $w'$ in the new network. Now we have $l(w')=\{1,2\}$. Therefore, we can always assume that there exists node $v_0$ with $|l(v_0)|=2$ such that $s \overset{\tilde{E}^c}{\longrightarrow} v_0$.

Consider the following sub-cases.

\textbf{Case 2a:} $l(v_0) = \{1,2\}$. Define a new labelling by
\[
l'(v) := \begin{cases}
\{1,2,3\} & \text{if}\; v \overset{\tilde{E}^c}{\longrightarrow} V_{12} \;\text{and}\; v \overset{\tilde{E}^c}{\longrightarrow} 3\\
\{3\} & \text{else if}\; v \overset{\tilde{E}^c}{\longrightarrow} 3\\
l(v) & \text{otherwise}.
\end{cases}\label{eqn:newlabel}
\]
where $v \overset{\tilde{E}^c}{\longrightarrow} V_{12}$ means there exists a node $v'\in V_{12}$ such that $v \overset{\tilde{E}^c}{\longrightarrow} v'$. By $l(v_0) = \{1,2\}$, we have $l(s) = \{1,2,3\}$. Similar to case 1, by considering the label $l'$ instead of $l$, we will design the coefficients $\beta_{12,1}$, $\beta_{12,2}$, $\beta_{123,12}$, $\beta_{123,3}$. Regardless of whether $l'(v_3)=\{1\}$, $\{2\}$ or $\{1,2\}$, the secrecy constraints are
\begin{align}
& |S_2| + \beta_{12,2} ( |S_{12}| + \beta_{123,12} |S_{123}| ) = 0,\\
& |S_3| + \beta_{123,3} |S_{123}| = 0.
\end{align}
The recovery requirement is
\begin{align}
&\beta_{12,1} + \beta_{12,2} = 1,\\
&(\beta_{12,1} + \beta_{12,2}) \beta_{123,12}+\beta_{123,3} = 1.
\end{align}
Solving these equations,
\begin{equation}
\left \{
\begin{array}{cl}
    \beta_{12,1} &= 1-\beta_{12,2}   \\
    \beta_{12,2} &= -\frac{|S_{2}|}{|S_{12}|+|S_{123}|+|S_{3}|}  \\
    \beta_{123,12} &= 1-\beta_{123,3}  \\
    \beta_{123,3} &= -\frac{|S_{3}|}{|S_{123}|}.
\end{array}
\right.
\end{equation}

\textbf{Case 2b:} $l(v_0) = \{2,3\}$. Define a new labelling by
\[
l'(v) := \begin{cases}
\{1,2,3\} & \text{if}\; v \overset{\tilde{E}^c}{\longrightarrow} V_{23} \;\text{and}\; v \overset{\tilde{E}^c}{\longrightarrow} 1\\
\{1\} & \text{else if}\; v \overset{\tilde{E}^c}{\longrightarrow} 1\\
l(v) & \text{otherwise}.
\end{cases}\label{eqn:newlabel}
\]
We have $l(s) = \{1,2,3\}$. By considering the label $l'$ instead of $l$, we will design the coefficients $\beta_{23,2}$, $\beta_{23,3}$, $\beta_{123,1}$, $\beta_{123,23}$. Regardless of whether $l'(v_3)=\{1\}$ or $\{2\}$, the secrecy constraints are
\begin{align}
& |S_2| + \beta_{23,2} ( |S_{23}| + \beta_{123,23} |S_{123}| ) = 0,\\
& |S_3| + \beta_{23,3} ( |S_{23}| + \beta_{123,23} |S_{123}| ) = 0.
\end{align}
The recovery requirement is
\begin{align}
&\beta_{23,2} + \beta_{23,3} = 1,\\
&(\beta_{23,2} + \beta_{23,3}) \beta_{123,23}+\beta_{123,1} = 1.
\end{align}
Solving these equations,
\begin{equation}
\left \{
\begin{array}{cl}
    \beta_{23,2} &= \frac{|S_2|}{|S_2|+|S_3|}  \\
     \beta_{23,3} &= \frac{|S_{3}|}{|S_{2}|+|S_3|}  \\
    \beta_{123,23} &= -\frac{|S_2|+|S_3|+|S_{23}|}{|S_{123}|}  \\
    \beta_{123,1} &= 1 - \beta_{123,23}.
\end{array}
\right.
\end{equation}

\medskip
\textbf{Case 2c:} $l(v_0) = \{1,3\}$. This is similar to the case $l(v_0) = \{1,2\}$. 

\medskip

\end{IEEEproof}

\fi

\section{Acknowledgement}

Cheuk Ting Li acknowledges support from the Direct Grant for Research,
The Chinese University of Hong Kong (Project ID: 4055133).

\ifshortver
\else
\appendix

\subsection{Proof of the Converse in Theorem~\ref{thm:converse}\label{subsec:pf_converse}}

Assume a source node $s \in S$ is weakly disconnected from the destination node $d$ after removing edges in $\mathrm{scl}(\tilde{E})$.
Our goal is to prove that $X_s$ (the source signal at $s$) is conditionally independent of the final output $Y_{d,\tilde{d}}$ at the destination, given $\{Y_{u,v}\}_{(u,v) \in \mathrm{scl}(\tilde{E})}$ (the signals along wiretapped edges). Due to the secrecy requirement  $\{Y_{u,v}\}_{(u,v) \in \mathrm{scl}(\tilde{E})} \perp \!\!\!\perp Y_{d,\tilde{d}}$, we have $X_s \perp \!\!\!\perp Y_{d,\tilde{d}}$, which is impossible since $Y_{d,\tilde{d}}$ is a sum of a set of source signals that contains $X_s$.

We will prove the desired conditional independence by constructing a Bayesian network~\cite{geiger1990identifying} on the signals along edges. Consider the directed line graph of $\tilde{G} := (V \cup \{0,\tilde{d}\},\, E \cup \{(0,s),(d,\tilde{d})\})$, where the set of vertices is $E \cup \{(0,s),(d,\tilde{d})\}$, and there is an edge from $(u_1,v_1)$ to $(u_2,v_2)$ if and only if $v_1=u_2$. Since the signal along edge $(u,v)$ only depends on the signals along incident edges at node $u$, the directed line graph gives a Bayesian network on the signals $Y_{u,v}$ along the edges. Also note that the Bayesian network contains $X_s = Y_{0,s}$ and the final output $Y_{d,\tilde{d}}$.

We now apply the concept of d-separation~\cite{geiger1990identifying}. If the desired conditional independence does not hold, then there exists an undirected path $(u_0,v_0)=(0,s),(u_1,v_1),\ldots,(u_{k-1},v_{k-1}),(u_k,v_k)=(d,\tilde{d})$ from $Y_{0,s}$ to $Y_{d,\tilde{d}}$ in the Bayesian network that is not blocked by $\mathrm{scl}(\tilde{E})$. If the connection at $(u_i,v_i)$ is serial to the right direction (i.e., the edges are $(u_{i-1},v_{i-1})\to (u_i,v_i) \to (u_{i-1},v_{i-1})$), then we have $v_{i-1}=u_i$, $v_{i}=u_{i+1}$. If the connection at $(u_i,v_i)$ is diverging (i.e., the edges are $(u_{i-1},v_{i-1})\leftarrow (u_i,v_i) \to (u_{i-1},v_{i-1})$), then we have $v_{i}=u_{i-1}=u_{i+1}$, and hence the nodes $v_{i-1},v_i,v_{i-1}$ are connected in the graph $\tilde{G}$ without going through $u_i$. If the connection at $(u_i,v_i)$ is converging (i.e., the edges are $(u_{i-1},v_{i-1})\to (u_i,v_i) \leftarrow (u_{i-1},v_{i-1})$), then we have $u_{i}=v_{i-1}=v_{i+1}$, and hence the nodes $u_{i-1},u_i,u_{i-1}$ are connected in $\tilde{G}$ without going through $v_i$. Hence, we can construct an undirected path from $s$ to $d$ in $\tilde{G}$ by finding the sequence of nodes visited in $(u_0,v_0),(u_1,v_1),\ldots,(u_k,v_k)$ after discarding each $(u_i,v_i)$ with a converging or diverging connection (keeping only serial connections). Since $s \in S$ is weakly disconnected from $d$ after removing edges in $\mathrm{scl}(\tilde{E})$, there must exist $(u_i,v_i) \in \mathrm{scl}(\tilde{E})$ with a serial connection, contradicting the d-separation requirement that the undirected path is not blocked by $\mathrm{scl}(\tilde{E})$. The result follows.

\fi

\bibliographystyle{ieeetr}
\bibliography{IEEEabrv,Refs.bib}



\end{document}